\documentclass{emulateapj}

\slugcomment{ApJ, in press}

\shorttitle{Predictions of planetary migration theory}
\shortauthors{Armitage}

\begin{document}

\title{Massive planet migration: Theoretical predictions \\ and comparison with observations}

\author{Philip J. Armitage\altaffilmark{1,2}}
\altaffiltext{1}{JILA, Campus Box 440, University of Colorado, Boulder CO 80309; 
pja@jilau1.colorado.edu}
\altaffiltext{2}{Department of Astrophysical and Planetary Sciences, University of Colorado, Boulder CO 80309}

\begin{abstract}
We quantify the utility of large radial velocity surveys for constraining theoretical models of Type~II 
migration and protoplanetary disk physics. We describe a theoretical model for the 
expected radial distribution of extrasolar planets that combines 
an analytic description of migration with an empirically calibrated disk 
model. The disk model includes viscous evolution and mass loss via photoevaporation. 
Comparing the predicted distribution to a uniformly selected subsample of planets from the 
Lick / Keck / AAT planet search programs, we find that a simple model in 
which planets form in the outer disk at a uniform rate, migrate  
inward according to a standard Type~II prescription, and become stranded when 
the gas disk is dispersed, is consistent with 
the radial distribution of planets for orbital radii $0.1~{\rm AU} \leq a < 2.5~{\rm AU}$ 
and planet masses $M_p > 1.65 \ M_J$. Some variant models are disfavored by 
existing data, but the significance is limited ($\sim$95\%) due to the small 
sample of planets suitable for statistical analysis. We show that the 
favored model predicts that the planetary mass function should be almost 
independent of orbital radius at distances where migration dominates the 
massive planet population. We also study how the radial distribution 
of planets depends upon the adopted disk model. We find that the distribution 
can constrain not only changes in the power-law index of the disk viscosity, but 
also sharp jumps in the efficiency of angular momentum transport that 
might occur at small radii. 
\end{abstract}


\section{Introduction}
Radial velocity and transit searches for extrasolar planets have detected in 
excess of 170 low-mass companions around nearby, mostly Solar-type, stars 
\citep{butler06}. These detections, which result from radial velocity surveys 
targeting a few~$\times 10^3$ stars, have allowed for an  
initial determination of the distribution of massive extrasolar 
planets with mass, orbital radius, eccentricity, 
and stellar metallicity \citep{santos04,marcy05,fischer05,santos05}. The 
statistical (and, hopefully, systematic) errors on these determinations  
will improve as ongoing surveys press to larger orbital radii and 
smaller planet masses. Substantially larger radial velocity surveys  
of $10^5$ to $10^6$ stars, with precision in the $\sim$10~ms$^{-1}$ range, 
are technically possible over the next decade \citep{ge06}. Given this rapid 
observational progress it is of interest to ask how much information -- 
about planet formation, planet migration, and the protoplanetary disk -- 
is retained in the statistical properties of extrasolar planets to be 
potentially tapped via an expansion of existing planet samples. Put more 
bluntly, is it worth obtaining much larger samples of planets with properties similar 
to those already known, or does the primary scientific interest for future 
surveys lie in exploring entirely new regimes of parameter space? 

At sufficiently small orbital radii, massive extrasolar planets very probably 
migrated inward from formation sites further out rather than forming 
in situ \citep{lin96,trilling98,bodenheimer00}. There remains some uncertainty in 
quantifying `sufficiently small', but it seems likely that massive 
planet formation is most common outside the snow line \citep{hayashi85}. 
Protoplanetary disk models clearly show that the radius of the snow line 
changes dramatically with time as the disk evolves \citep{garaud07}, so 
to quote a single radius is potentially misleading. However, for a Solar 
type star, the apparent presence of hydrated minerals in Solar System 
asteroids allows an empirical determination of the location of the 
snow line at a radius of around 2.7~AU \citep{morbidelli00}. This 
suggests that most of the extrasolar planets currently known, which orbit within  
a few~AU of their host stars, derive their properties largely via 
migration. For massive planets, the appropriate regime of migration 
is thought theoretically to be the Type~II regime --- which involves migration within a gap in 
the protoplanetary disk \citep{goldreich80,lin86} --- rather than the 
gap-less Type~I regime appropriate to Earth mass planets and giant planet cores 
\citep{ward96}. Since direct observational evidence of migration is 
currently lacking, the only tests possible of this theory come 
from statistical comparison with the observed properties of extrasolar 
planetary systems. Indeed, prior work along these lines by \cite{armitage02}, 
\cite{trilling02} and \citet{ida04} has shown that the distribution 
of observed planets in orbital radius (and, in the case of the 
\cite{ida04} study, planetary mass) is broadly consistent with theoretical 
expectations based on disk migration within an evolving protoplanetary 
disk.

In this paper, we develop more refined predictions for the radial distribution 
of massive planets based on a simple analytic model for Type~II 
migration. Our main goal is to determine, at least in principle, what might be learned 
from comparisons of large planet samples with theoretical models. In \S2 we describe the 
adopted model for the protoplanetary disk, and how migration 
of massive planets within the disk is treated. In \S3 we 
compute the predicted distribution of planets in orbital 
radius. We compare the predictions to the observed distribution 
of planets in the \cite{fischer05} sample, which has previously 
been used to study the dependence of planet frequency on host 
metallicity. This sample has a clearly 
specified selection limit, which allows for a 
reliable statistical comparison between the model and observations. 
In \S4 and \S5 we investigate the extent to which migration 
leads to a radial variation in the exoplanet mass function, 
and how sensitive it is to structure within the protoplanetary 
disk. These Sections are primarily forward-looking, since 
existing data are too limited to support or refute the 
theoretical model. We conclude in \S6 with some discussion of 
the results.

\section{Type II migration within the protoplanetary disk}
We consider a model in which massive planets form within an 
evolving protoplanetary disk at radii beyond the snow line, 
and migrate inward within a gap (`Type~II' orbital migration). 
Migration slows and eventually ceases as the gas disk is dispersed. We assume that 
photoevaporation causes disk dispersal. The resulting distribution 
of massive planets in orbital radius then depends upon the 
disk model (which is reasonably tightly constrained by observations); 
the migration rate (which is known reasonably well theoretically); 
and the rate of planet formation in the disk as a function of 
time. The latter can in principle be predicted from a model of massive 
planet formation, but here we treat it as a free function.

The surface density $\Sigma$ of a protoplanetary disk that 
evolves under the combined action of an effective kinematic viscosity 
$\nu$, and mass loss per unit area $\dot{\Sigma}_{\rm wind} (r)$, is described 
by \citep{pringle81},
\begin{equation} 
 \frac{\partial \Sigma}{\partial t} = \frac{3}{r} 
 \frac{\partial}{\partial r} \left[ 
 r^{1/2} \frac{\partial}{\partial r} \left( \nu 
 \Sigma r^{1/2} \right) \right] - \dot{\Sigma}_{\rm wind} (r),
\label{eq_diffusion}
\end{equation} 
provided that the mass lost in the wind has the same specific 
angular momentum as the disk at the launch point. Approximating 
the angular momentum transport in the disk as a time-independent 
kinematic viscosity that is a power-law in radius,
\begin{equation} 
 \nu = \nu_0 r^\gamma
\label{eq_viscous} 
\end{equation}
equation (\ref{eq_diffusion}) admits a compact self-similar solution 
if the mass loss rate is negligible \citep{lyndenbell74,hartmann98}. 
In this solution the surface density 
evolves according to,
\begin{equation} 
 \Sigma (R,T) = \frac{C}{3 \pi \nu_{\rm scale} R^\gamma} 
 T^{-(5/2 - \gamma)/(2-\gamma)} 
 \exp[{-R^{(2-\gamma)} / T}]
\label{eq_similar} 
\end{equation}  
where the scaled variables $R$ and $T$ are defined via a 
fiducial radius $r_{\rm scale}$,
\begin{eqnarray}
 R & \equiv & \frac{r}{r_{\rm scale}} \\
 \nu_{\rm scale} & \equiv & \nu(r_{\rm scale}) \\
 T & \equiv & \frac{t}{t_{\rm scale}} + 1 \\
 t_{\rm scale} & = & \frac{1}{3(2-\gamma)^2} 
 \frac{r_{\rm scale}^2}{\nu_{\rm scale}}.
\end{eqnarray}
The constant $C$ is the mass accretion rate at $t=0$ as 
$r \rightarrow 0$. Other quantities, such as the mass accretion rate 
$\dot{M}$ and radial velocity $v_r$ in the disk, can be derived straightforwardly 
using these expressions.

\begin{figure}
\plotone{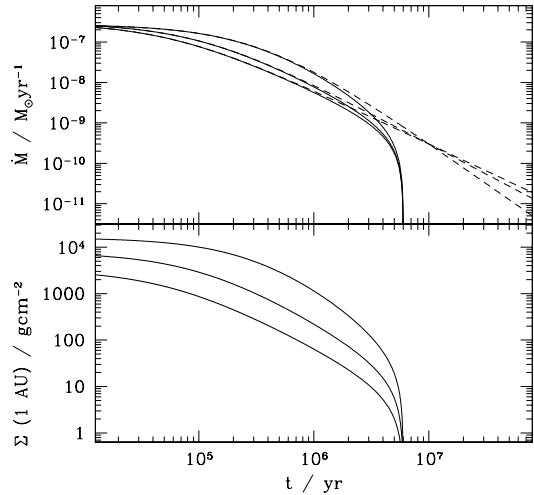}
\caption{The time dependence of the accretion rate on to the central star plotted for 
different disk models. The three dashed curves in the upper panel show the predicted evolution in 
self-similar models with different power-law exponents for the disk viscosity: 
$\gamma=0.5$, $\gamma=1.0$ and $\gamma=1.5$. The solid curves show the 
result including photoevaporation, which is here modeled using the solution 
of \cite{ruden04} assuming a disk dispersal time of 6~Myr. The analytic 
cut-off is derived assuming $\gamma=1.0$, but should be a good approximation 
for general power-law viscosity profiles. The lower panel shows the surface 
density evolution at a disk radius of 1~AU (the $\gamma=1.5$ model is the 
uppermost curve).}
\label{f1}
\end{figure}
 
We constrain the parameters of the similarity solution 
(equation \ref{eq_similar}) to be consistent with the observational 
analysis of the time dependence of T~Tauri accretion rates presented 
by \cite{muzerolle00}. We fix all models to have an initial accretion 
rate $\dot{M} (t=0) = 3 \times 10^{-7} \ M_\odot {\rm yr}^{-1}$, 
which decays to $\dot{M} = 3 \times 10^{-10} \ M_\odot {\rm yr}^{-1}$ 
in $10^7$~yr if the disk lives that long before being dispersed. We also 
fix $r_{\rm scale} = 10 \ {\rm AU}$. Table~1 shows the 
values of viscosity normalization constant $\nu_0$, 
and initial disk mass $M_{\rm disk}$, that meet these constraints 
for different values of power-law index $\gamma$. Not surprisingly, 
since the basic observational constraints have been known for 
some time, the disk models we favor have parameters similar to 
those considered in many previous studies \citep{hartmann98,armitage03,
alexander06}. Figure~\ref{f1} 
shows the resulting evolution of the accretion rate for these  
models.
 
\begin{table*}
\center{
\begin{tabular}{ccc}
\hline \hline
Power law index of viscosity $\gamma$ & Viscosity normalization $\nu_0$ & Initial disk mass $M_{\rm disk}$ \\ \hline
0.5 & $1.52 \times 10^8 \ {\rm cm}^{3/2}{\rm s}^{-1}$ & $0.057 \ M_\odot$ \\
1.0 & $15.62 \ {\rm cm}{\rm s}^{-1}$ & $0.051 \ M_\odot$ \\
1.5 & $1.58 \times 10^{-6} \ {\rm cm}^{1/2}{\rm s}^{-1}$ & $0.058 \ M_\odot$ \\ 
\hline
\label{table1}
\end{tabular}}
\caption{Properties of self-similar disk models that meet the 
observational constraints \\ on the evolution of the stellar 
accretion rate discussed in the text}
\end{table*} 
 
The power-law decline in the late-time accretion rate implied by the 
similarity solution does not yield the sharp transition between 
accreting Classical T~Tauri stars and non-accreting Weak-Lined T~Tauri 
stars that is observed \citep{simon95,wolk96}. It is plausible that this 
transition is driven by photoevaporation \citep{bally82} from the outer disk, 
which acts to starve the inner disk (thereby allowing it to drain viscously 
onto the star on a short time scale) once the accretion rate 
becomes comparable to the wind mass loss rate \citep{clarke01}. 
If photoevaporation is  
driven by irradiation from the central star, then the 
simplest analytic models predict that the mass loss 
rate per unit area scales as \citep{hollenbach94},
\begin{eqnarray}
 \dot{\Sigma}_{\rm wind} \propto r^{-5/2}
\label{eq_massloss} 
\end{eqnarray}
exterior to some critical radius $r_{\rm in}$, with zero 
mass loss from smaller radii. The critical radius is given 
to order of magnitude by the radius where the sound speed 
in photoionized gas ($T \simeq 10^4$) first exceeds the local 
escape speed. This is a few AU for a Solar mass star. The 
normalization of the mass loss rate depends upon the square 
root of the ionizing flux $\Phi$, which is hard to measure 
accurately but whose value can be constrained observationally 
\citep{alexander05,pascucci07}. Much 
more detailed numerical models of photoevaporation are now 
available \citep{font04,alexander06}, but the additional 
complexity they involve is not warranted for this application.
Accordingly, we assume a time-independent mass loss rate of the form given 
by equation (\ref{eq_massloss}), with $r_{\rm in} = 5 \ {\rm AU}$, and 
zero mass loss at smaller radii.

Once photoevaporation is included (in this simplified form), it is still 
possible to derive a Green's function solution to equation (\ref{eq_diffusion}) 
\citep{ruden04}, but there is no compact form for the evolution of 
$\Sigma(r,t)$ analogous to equation (\ref{eq_similar}). At late times 
(i.e. after the disk has evolved for several viscous times) and small 
radii ($r < r_{\rm in}$), however, it is possible to derive an 
expression for the reduction in the inner accretion rate and 
surface density caused by the wind. Defining $t_0$ as the time 
at which the inner accretion rate falls to zero as a consequence of 
the mass loss from the outer disk, \cite{ruden04} finds that the 
time dependence of the accretion rate at small radii follows,
\begin{equation}
 \dot{M} = \left[ 1 - \left( \frac{t}{t_0} \right)^{3/2} \right]
 \dot{M}_{\rm self-similar}, 
\end{equation} 
where $\dot{M}_{\rm self-similar}$ is the accretion rate evolution 
predicted by the self-similar model in the absence of any mass loss. 
The suppression of accretion implied by the term in parenthesis is 
derived for $\gamma=1$, but should also be approximately valid for 
other values of the power law viscosity index. Using numerical 
solutions to equation (\ref{eq_diffusion}), we have verified that 
the \cite{ruden04} formula provides a good description of the 
evolution of the accretion rate and inner surface density during 
the transition, and hence we apply the same cutoff to generate 
the surface density evolution including wind loss $\Sigma(r,t)$ from 
the self-similar prediction (equation~\ref{eq_similar}). Based on 
observations of disk frequency in young clusters \citep{haisch01}, 
we take $t_0 = 6 \ {\rm Myr}$, which yields an 
accretion rate evolution shown as the solid curves in Figure~\ref{f1}.
As has been emphasized previously \citep{clarke01,armitage03},   
to a good approximation the disk evolution proceeds as if there is no mass 
loss while $\dot{M} \gg \dot{M}_{\rm wind}$, and is then dispersed rapidly 
once mass loss becomes significant. Figure~\ref{f1} also shows the 
evolution of the inner (1~AU) disk surface density. Irrespective of 
the value of $\gamma$, all of the disk models considered have an 
{\em initial} surface density that is substantially in excess of 
the minimum mass Solar Nebula reference value \citep{weidenschilling77}, 
but this drops rapidly as the disk evolves. By the time that massive 
planets migrate through this region (after several Myr of evolution) 
the surface density has fallen by at least an order of magnitude. The 
import of this is that the local disk mass during migration is 
typically smaller than the mass of giant planets, a regime which 
leads to significantly slower inspiral.

Sufficiently massive planets are able to open a gap within the gaseous 
protoplanetary disk and, having opened a gap, migrate inward\footnote{Here, 
we consider exclusively planets that form close enough in that the sense 
of gas flow and migration is inward --- planets that form further out 
may instead absorb the angular momentum of the inner disk and move 
outward \citep{veras04,martin07}, or become stranded outside the annular hole 
that is predicted to form as photoevaporation proceeds \citep{clarke01,matsuyama03}.} 
in lockstep with the gas via Type~II migration \citep{lin86}. For a planet at 
orbital radius $r_p$, the gravitational torques exerted by the planet on the 
surrounding disk are strong enough to maintain a gap provided that 
the mass ratio $q = M_p / M_*$ satisfies \citep{takeuchi96},
\begin{equation}
 q \gtrsim \left( \frac{c_s}{r_p \Omega_p} \right)^2 \alpha^{1/2}
\end{equation}
where $c_s$ and $\Omega_p$ are the sound speed and angular velocity 
in the disk at the radius of the planet, and $\alpha$ is the 
dimensionless Shakura-Sunyaev viscosity parameter \citep{shakura73}. 
Noting that the thickness of the disk $h \simeq c_s / \Omega_p$, 
we find that the $\gamma=1$ disk model specified in Table~1 
would correspond to an equivalent $\alpha \simeq 5 \times 10^{-3}$ 
(matching at 5~AU around a Solar mass star assuming that the 
disk has $h/r = 0.05$). The viscous gap opening criteria is then 
satisfied for planets of mass exceeding about 0.2~$M_J$. The 
thermal gap opening condition is satisfied at a similar mass 
\citep{bate03}. In this paper, we exclusively consider planets 
with masses of $M_J$ and above, which should accordingly be safely 
in the gap-opening regime of parameter space. 

\begin{figure}
\plotone{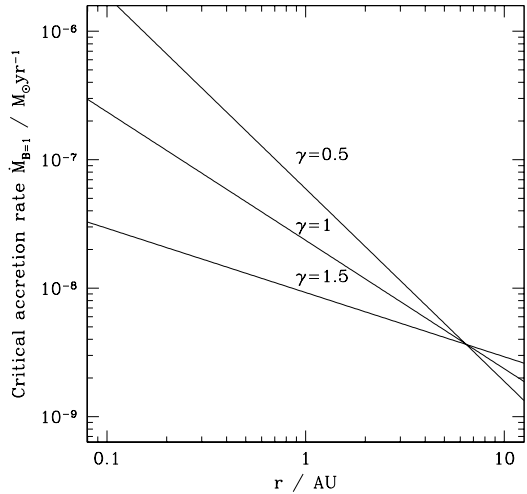}
\caption{The critical accretion rate above which a 1~$M_J$ planet migrates as a 
test particle is plotted as a function of radius for the different disk models.}
\label{f2}
\end{figure}

\begin{figure}
\plotone{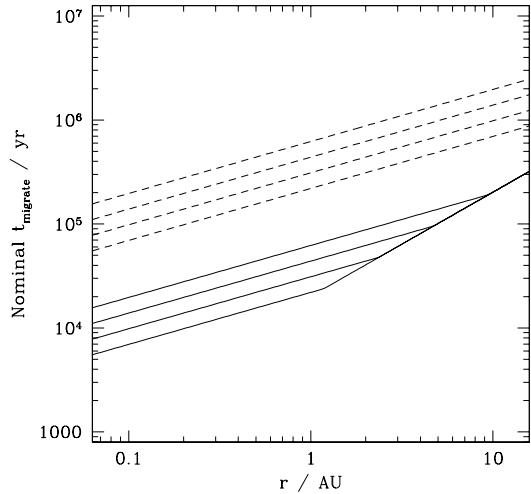}
\caption{The predicted migration time scale as a function of radius in the 
$\gamma=1$ disk model. The solid lines show the migration time scale for 
planets of masses 0.5~$M_J$, 1~$M_J$, 2~$M_J$ and 4~$M_J$ when the disk 
accretion rate is $10^{-8} \ M_\odot {\rm yr}^{-1}$ (more massive planets 
have longer migration time scales at small orbital radii). The dashed lines 
show results for the same mass planets in a disk with $\dot{M} = 10^{-10} 
\ M_\odot {\rm yr}^{-1}$.}
\label{f3}
\end{figure}

To compute the Type~II migration rate within the disk model 
specified above, we draw on the results of \cite{syer95}. 
These authors noted that the rate of migration, which is 
sometimes assumed to equal the radial velocity that gas in 
the disk {\em would have in the absence of a planet}, is 
suppressed once the planet mass exceeds a local estimate 
of the disk mass. Specifically, they defined a measure of 
the importance of the disk relative to the planet,
\begin{equation}
 B \equiv \frac{4 \pi \Sigma r_p^2}{M_p}
\end{equation}
which is small if the planet dominates the angular momentum 
budget of the planet + disk system, and large otherwise. The 
Type~II migration rate is then given for $B \geq 1$ (the disk 
dominated case) by,
\begin{equation}
 \dot{r}_p = v_r 
\label{eq_migrate1} 
\end{equation}
while for $B < 1$ (the planet dominated limit),
\begin{equation}
 \dot{r}_p = B^{1/2} v_r
\label{eq_migrate2}
\end{equation}
where we have assumed (consistent with equation~\ref{eq_viscous}) 
that the efficiency of angular momentum transport is independent 
of the surface density. For our disk models, Figure~\ref{f2} shows the critical accretion rate 
below which $B < 1$ for a Jupiter mass planet. In all these models, slowdown 
of the migration rate occurs due to the inertia of the planet 
for accretion rates substantially greater than those at 
which photoevaporation starts to influence the disk evolution. 
The effect on the nominal migration time scale,
\begin{equation} 
 t_{\rm migrate} \equiv \frac{r_p}{\left| \dot{r}_p \right|}
\end{equation}
is shown in Figure~\ref{f3}. The $B < 1$ limit is applicable 
at all radii of interest in a disk with $\dot{M} = 10^{-10} \ 
M_\odot {\rm yr}^{-1}$, and within radii less than a few 
AU (depending on the planet mass) in a disk with 
$\dot{M} = 10^{-8} \ M_\odot {\rm yr}^{-1}$. 

Once planets exceed the critical gap opening mass by a significant 
factor, the rate of accretion onto the planet across the gap 
drops rapidly \citep{lubow99}. Our goal is to compare theoretical 
predictions for the planet distribution with a complete subsample 
of extrasolar planets, which requires a cut at approximately 
1.5~$M_J$. Most of the planets that survive this cut have masses 
substantially greater than the gap-opening mass, so {\em as a 
first approximation} it seems reasonable to ignore the possibility 
of mass accretion during Type~II migration, and assume that 
negligible mass is accreted across the range of radii (interior 
to 2.5~AU) over which we make the comparison. The results of 
\cite{bate03}, which show that planets can grow rapidly to 
masses beyond that of Jupiter while suffering little migration, 
support this approximation. It is then easy, using 
equations (\ref{eq_migrate1}) and (\ref{eq_migrate2}), to calculate 
the final (following disk dispersal) orbital radius of a planet 
that forms with mass $M_p$ at radius $a_{\rm form}$ and time $t_{\rm form}$ 
in one of the disk models specified in Table~1. The resulting mapping 
between the formation conditions and the final state is the basic 
input needed to make a prediction of the resulting planet distribution.

The above model represents our attempt to define a `best-guess' description 
of Type~II migration within the protoplanetary disk. One may note that 
in the planet-dominated $B < 1$ limit, the predicted rate of Type~II 
migration is reduced by the square-root of $B$ rather than the full ratio 
of the local disk mass to the planet mass. We will refer to this as 
{\em partial suppression} of the migration rate. The rate is not 
fully suppressed because gas accumulates close to the tidal barrier 
as the disk / planet system evolves, increasing the torque beyond 
the value that would occur in an unperturbed disk \citep{pringle91}. 
In practice, however, some gas may overflow the gap to be either 
accreted by the planet or to flow into an inner disk interior 
to the gap \citep{lubow06}. Mindful of this, we have also 
computed models in which the migration rate in the planet dominated 
regime is {\em fully suppressed}, so that,
\begin{equation}
 \dot{r}_p = B v_r.
\end{equation}
The difference between the partially and fully suppressed calculations 
provides an indication as to how sensitive the results are to uncertainties 
in the treatment of massive planet migration.

\section{Predicted distribution of planets in orbital radius}
\begin{figure}
\plotone{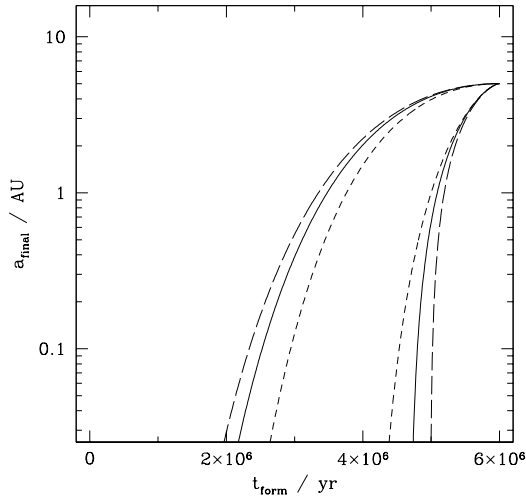}
\caption{Predicted final radii for 1~$M_J$ planets as a function of the time at 
which they formed at 5~AU in the protoplanetary disk. The three closely spaced 
curves on the right-hand-side of the Figure show results for disks with 
$\gamma=0.5$ (long-dashed lines), $\gamma=1$ (solid lines) and $\gamma=1.5$ 
(short-dashed lines) respectively, assuming that migration is suppressed at 
small radii according to the \cite{syer95} model. The left-hand curves show 
results if, instead, migration is fully suppressed by the ratio of the local 
disk mass to the planet mass. Differences between migration treatments are 
evidently more significant that differences between the disk models. For the 
favored models in which migration is partially suppressed, the survival 
time of planets in the disk is in the range of 10-20\% of the disk lifetime.}
\label{f4}
\end{figure}

Using the analytic disk evolution and migration models described in 
\S2, we compute the radius $a_{\rm final}$ at which massive planets become 
stranded as a function of the time $t_{\rm form}$ at which they form 
within the protoplanetary disk. The parameters of the model are the planet mass, 
the power-law index of the disk viscosity, the formation radius, 
and whether the migration rate is partially or fully suppressed. 
Illustrative results for a formation radius of 5~AU and a planet mass 
of 1~$M_J$ are shown in Figure~\ref{f4}. As is well known, planets 
that form at earlier epochs migrate to smaller final radii, and  
as the orbital radii decrease the window of allowed formation 
times also narrows \citep{trilling02,armitage02}. For the favored model, in which migration is only 
partially suppressed, all surviving planets must form during the 
last 1-1.5~Myr of the disk lifetime. This is not so short a window as to 
imply that the planet formation process must be worryingly lossy, but it 
does imply that the final masses of giant planets might well reflect details of the disk 
dispersal process \citep{shu93}. Planets can form and survive across 
a larger fraction of the disk lifetime if instead migration is 
fully suppressed. There are also significant differences in the 
outcome that depend upon the adopted disk model. These arise 
because of the differing profiles of the radial velocity as a 
function of radius, but they are less significant than differences in the 
treatment of migration.

\begin{figure}
\plotone{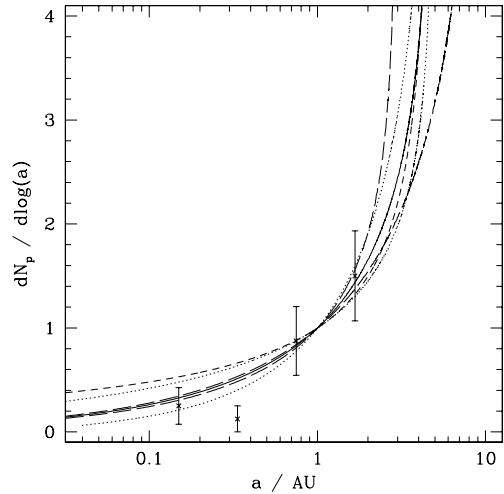}
\caption{The predicted distribution of massive planets as a function of radius. 
All the curves show results for Jupiter mass planets (note that the mass dependence 
is sufficiently weak that essentially identical results apply also for higher 
masses). The solid curve shows the 
results for the favored model, in which planets form at 5~AU at a constant 
rate in a $\gamma=1$ disk, and migration is partially suppressed at small 
radii. The two dotted curves show the effect of varying the disk viscosity 
to $\gamma=0.5$ or $\gamma=1.5$. The long-dashed curves show the effect of 
assuming that planets form at 3~AU (the curve furthest to the left at 
2.5~AU) or 10~AU rather than 5~AU. The short-dashed curve is  
the predicted distribution in the case in which migration is fully suppressed. 
The data points (shown with approximate errors) show the number of observed 
extrasolar planets with $1.65 M_J < M_p \sin(i) < 10 M_J$ in a uniform and 
complete sample of planets constructed from \cite{fischer05}.}
\label{f5}
\end{figure}

To translate these results into a prediction for the orbital distribution 
of massive planets, we note that,
\begin{equation}
 \frac{{\rm d}N_p}{{\rm d}a} = \frac{{\rm d}t_{\rm form}}{{\rm d}a_{\rm final}} \times 
 \frac{{\rm d}N_p}{{\rm d}t_{\rm form}}
\end{equation}
where ${{\rm d}N_p}/{{\rm d}t_{\rm form}}$ is the rate at which planets 
of a given mass form in the outer disk. The simplest assumption is that 
this rate is a constant, or at least can be approximated as such, over 
the interval of time near the end of the disk lifetime during which 
planets can form and survive migration without being swept into the star. 
Making this assumption, we compute and plot in Figure~\ref{f5} 
${{\rm d}N_p}/{{\rm d}\log a}$ for a variety of models, which differ 
in the assumed formation radius, disk model, and migration treatment. 
Numerical values for three of the these models, which assume 
$a_{\rm form} = 5 {\rm AU}$, partial suppression of the migration 
velocity, and $\gamma=0.5$, $\gamma=1$ or $\gamma=1.5$, are tabulated 
in Table~2. These three cases roughly bracket the range of predicted 
outcomes for all of the models that we have considered. 

\begin{table*}
\center{
\begin{tabular}{cccc}
\hline \hline
$\log(a / {\rm AU})$ & ${\rm d}N (\gamma=0.5) / {\rm d}\log(a) $ & ${\rm d}N (\gamma=1.0) / {\rm d}\log(a) $ &
${\rm d}N (\gamma=1.5) / {\rm
d}\log(a) $ \\ 
\hline
-1.0 & 0.152 & 0.263 & 0.420 \\
-0.9 & 0.184 & 0.298 & 0.423 \\
-0.8 & 0.221 & 0.338 & 0.489 \\
-0.7 & 0.266 & 0.384 & 0.529 \\
-0.6 & 0.320 & 0.436 & 0.573 \\
-0.5 & 0.385 & 0.497 & 0.623 \\
-0.4 & 0.463 & 0.567 & 0.679 \\
-0.3 & 0.559 & 0.650 & 0.743 \\
-0.2 & 0.676 & 0.747 & 0.816 \\
-0.1 & 0.820 & 0.863 & 0.899 \\
0.0 & 1.000 & 1.000 & 1.000 \\
0.1 & 1.226 & 1.170 & 1.117 \\
0.2 & 1.518 & 1.380 & 1.261 \\
0.3 & 1.912 & 1.656 & 1.448 \\
0.4 & 2.458 & 2.035 & 1.705 \\
0.5 & 3.283 & 2.592 & 2.080 \\
\hline
\label{table2}
\end{tabular}}
\caption{The predicted number of planets per logarithmic interval in 
semi-major axis for models in which planet formation occurs at 5~AU and 
the migration rate is partially suppressed. Results for three different 
values of $\gamma$ are quoted. These results approximately bracket the 
shallowest to steepest curves obtained for the models shown in Figure~\ref{f5}. 
Results are quoted for 1 Jupiter mass planets, but the mass dependence is 
weak -- much weaker than the differences between the different $\gamma$ models. 
The numbers have been normalized to unity at 1~AU.}
\end{table*}

Although all of the curves have the same general form --- relatively 
few planets are predicted to be stranded at very small orbital radii, 
with the number increasingly rapidly as $a$ increases --- there are 
significant differences which may in principle leave signatures in the 
observed distribution of extrasolar planets. Most importantly, the 
predicted number of planets per logarithmic interval in semi-major 
axis rises sharply as the formation radius is approached. A model in 
which planets are assumed to form typically at 10~AU rather than 
5~AU results in a much smaller predicted number of planets at 4~AU, 
whereas assuming a typical formation radius of 3~AU (i.e. immediately 
outside the snow line) leads to a predicted pile-up of planets as that 
radius is approached. 
Observational detection of a rapid rise in the number of planets at a 
particular orbital radius would then be a signature of a preferred 
orbital radius for giant planet formation. Also easily detectable 
are the differences between the partially and fully suppressed 
migration models. If migration is fully suppressed, then the 
number of planets (expressed as ${{\rm d}N_p}/{{\rm d}\log a}$) at 
0.1~AU is predicted to be around half the number at 1~AU, whereas 
for partial suppression this ratio is around 0.25. This result is 
easily understood -- differences in the treatment of migration are 
most important at those small radii, interior to 1~AU, where the disk 
mass is lowest. 

\subsection{Sensitivity to the disk dispersal mechanism}
Our parameterization of disk evolution sweeps much that is poorly 
known about photoevaporation into a single parameter -- the disk 
dispersal time $t_0$. Physically, rapid dispersal (small values of 
$t_0$ for a given set of disk initial conditions) will occur 
for large values of the ionizing flux $\Phi$, which will drive a 
stronger photoevaporative flow. Quantitatively, the total mass 
loss rate from the disk scales as \citep{hollenbach94},
\begin{equation}
 \dot{M}_{\rm wind} \simeq 1.5 \times 10^{-10} 
 \left( \frac{\Phi}{10^{41} \ {\rm s}} \right)^{1/2} 
 \left( \frac{M_*}{M_\odot} \right) ^{1/2} M_\odot {\rm yr}^{-1}
\end{equation}
where the prefactor has been adjusted from the analytic 
value to better match the numerical results of \cite{font04}, and the 
fiducial ionizing flux has been chosen to be consistent with observational 
estimates \citep{alexander05,pascucci07}, which are, however, subject 
to substantial uncertainties.

\begin{figure}
\plotone{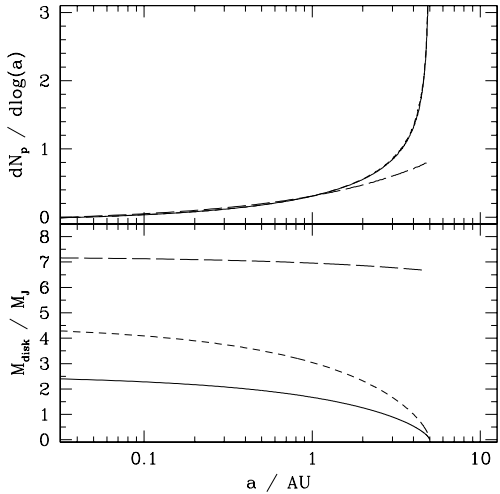}
\caption{The predicted distribution of massive planets as a function of radius 
(upper panel) is plotted together with the disk mass evaluated at the formation 
epoch (lower panel). The solid curve shows the results for the favored model, in which planets 
form at 5~AU at a constant rate in a $\gamma=1$ disk, and migration is partially 
suppressed at small radii. The short-dashed curve -- almost indistinguishable in the 
upper panel -- shows the effect of reducing the disk dispersal time $t_0$ from 
6~Myr to 3~Myr. The long-dashed curve shows the results of assuming an instantaneous 
dispersal of the disk at 6~Myr.}
\label{f5b}
\end{figure}

With our disk models, a wind mass loss rate in the range between 
$10^{-10} \ M_\odot {\rm yr}^{-1}$ and $10^{-9} \ M_\odot {\rm yr}^{-1}$ 
yields the dispersal time of 6~Myr that is our baseline assumption. 
Our fiducial parameters are therefore consistent with standard 
photoevaporation models. A consequence of this, however, is that 
the {\em disk mass} at the epoch when surviving planets form is 
quite small. Figure~\ref{f5b} shows the gas disk mass at the moment 
planets form as a function of the final planet semi-major axis. For 
the models we have considered, the typical disk masses are only a 
few Jupiter masses if migration is partially suppressed (in the 
fully suppressed case, on the other hand, surviving planets form 
substantially earlier when there is plenty of disk gas remaining). 
These small masses directly reflect the efficiency of the migration process -- 
relatively modest amounts of gas are able to drive substantial 
migration. However, they do pose a possible consistency problem: 
planets that form in very low mass disks evidently perturb the 
disk structure substantially, and this might affect the final 
planetary distribution.

To explore this possibility, we considered two options. First, 
within the formalism developed here, we have investigated the 
effect of reducing $t_0$ so that surviving planets form earlier 
when the disk mass is higher. As shown in Figure~\ref{f5b}, 
a model computed with $t_0 = 3 \ {\rm Myr}$ yields an almost 
identical planet distribution. Second, we have compared the 
results for $\gamma=1.5$ with the numerical models presented 
in \cite{armitage02}. In the numerical models, planet masses 
are restricted to be no larger than a local estimate of the 
disk mass, and the feedback of a massive planet on the structure 
of a very low mass disk is explicitly followed. Reasonably 
good agreement is obtained between the numerical and analytic 
schemes, with the magnitude of the differences being comparable 
to the differences between the curves plotted in Figure~\ref{f5}.

A related question is how important the nature of the disk 
dispersal mechanism is for the resulting planet distribution. 
Physically, a model in which there is {\em no} mass loss from 
the disk exterior to the planet (either via a wind, or via 
accretion on to the planet) fails to yield any surviving planets -- 
ultimately the disk, no matter how small, absorbs the orbital angular momentum of the 
planet and drives it to small radius\footnote{Strictly, this 
is only true if the inner disk is subject to a zero-torque 
boundary condition. A steady-state solution is possible if 
there is instead angular momentum injection at small radii, 
for example due to the interaction between the disk and 
the stellar magnetosphere \citep{armitage96}.}. However, 
many other models in which the disk is dispersed rapidly 
do yield sensible distributions, so it is of interest to 
assess how sensitive the planet distribution is to the 
specifics of disk dispersal. To gauge this, we have considered 
an extreme model in which the disk evolves viscously for 
6~Myr without any mass loss, and is then instantaneously 
dispersed. Figure~\ref{f5b} depicts the resulting planet 
distribution. At small radii (within about 1.5~AU) this 
model tracks the fiducial photoevaporative case closely, 
but further out the instantaneous dispersal model yields 
a much flatter distribution. This reflects the fact that 
in the case of photoevaporation, migration of the last 
planets to form is limited by the rapid drop in the disk 
surface density, and these planets pile up at radii 
relatively close to their formation sites. We conclude that 
the innermost part of the extrasolar planet distribution 
(within roughly an AU) ought to be largely independent of 
how the disk is dispersed, but that the distribution 
further out does depend on whether photoevaporation or some 
other mechanism is at work.

\subsection{Effect of the dispersion in disk properties}
The disk models used in this paper have been adjusted to 
approximately reproduce the mean lifetimes and accretion rates 
of observed disks \citep{haisch01,hartmann98}. This procedure 
ignores the fact that, observationally, there is a large 
dispersion in the accretion rate at a given age. This 
dispersion exceeds that expected from measurement uncertainties in 
accretion rates and stellar ages.

The origin of the intrinsic dispersion in disk properties 
as a function of system age is not known, and hence it is 
impossible to make a blanket statement as to whether consideration 
of a mean model is reasonable or not. We can, however, distinguish 
some possibilities. One possibility \citep{armitage03} is that the dispersion 
in disk properties arises from a dispersion in disk initial 
conditions (disk mass, disk angular momentum). Since the 
planet distribution depends only upon the disk evolution 
close to the transition epoch (independent of the absolute 
timing), in this scenario the ensemble distribution averaged over the population 
would be expected to be the same as the mean model we have 
computed. Of course, giant planet formation would be less 
probable in short-lived disks with low initial surface 
densities \citep{pollack96}, so in practice most planets 
would form in those disks with higher initial surface 
densities at radii of a few AU.

A second possibility is that the observed dispersion in 
accretion rates arises from changes in individual accretion 
rates on timescales intermediate between the disk lifetime 
and observable timescales. Apart from thermal instabilities 
\citep{bell94}, which are only likely to be relevant for 
initial accretion rates higher than those we have considered here, 
no obvious candidate instabilities that might yield large fluctuations 
in $\dot{M}$ are known, but it remains possible that $\dot{M} (r)$ 
might vary rapidly. If such variability extended across the 
radial range considered here, it is likely that the migration 
history of planets would be altered.

\subsection{Comparison with a uniformly selected dataset}
We compare the models to a subsample of extrasolar planets detected 
by the Keck / Lick / AAT planet search program. Initially, we proceed 
rather conservatively, and define 
a subsample that is as complete and unbiased as 
possible over specified intervals in planet mass and orbital radii. 
Our procedure is as follows:
\begin{enumerate}
\item
We start with a sample of 850 stars of FGK spectral types, targeted 
by the Lick / Keck / AAT planet search program, for which 10 or 
more radial velocity measurements over a period of at least 4 
years are available. This sample is listed in Table~3 of 
\cite{fischer05}. For these stars, hypothetical planets that 
yield a radial velocity semi-amplitude $K > 30 \ {\rm ms}^{-1}$ 
have nearly uniform detectability provided that the orbital 
period is less than 4~years. The orbital period restriction 
corresponds to a maximum semi-major axis of 2.5~AU, assuming 
Solar mass hosts. From this sample, \cite{fischer05} list 46 
stars which host known extrasolar planets. In some cases, these are 
multiple systems.
\item
In many cases, additional radial velocity data is available beyond 
that used by \cite{fischer05}. We therefore update the orbital 
elements quoted by \cite{fischer05} to match those reported by 
\cite{butler06}. Typically the changes to the derived masses and 
semi-major axes are rather modest (here we ignore eccentricity, 
which in some cases is more substantially altered). Including the 
multiple planets in some systems, this yields a sample of 59 planets.
\item
Finally, we define a subsample that includes only planets that are 
massive enough to be detectable across the entire range of orbital radii 
for which the survey is complete. At 2.5~AU, $K = 30 \ {\rm ms}^{-1}$ 
corresponds to a planet mass of $1.65 \ M_J$, so we discard planets 
with smaller masses. We also cut the sample at the high mass end, 
somewhat arbitrarily, at $10 M_J$. In terms of radial extent, we 
include those planets with $0.1 \ {\rm AU} < a < 2.5 \ {\rm AU}$. 
This excludes planets with very tight orbits whose dynamics may have 
been influenced by tidal effects and / or penetration into the 
protostellar magnetosphere.
\end{enumerate}
It is somewhat striking --- given the large number of extrasolar 
planets now known --- how little of the publicly available data 
can be used for the sort of statistical comparison we are attempting 
here. Demanding both completeness (which restricts the stellar sample 
used), and lack of mass bias (which restricts the minimum mass), we are 
left with only 22 massive planets. Obviously this small sample size 
limits the statistical power to constrain theoretical models. However, 
the sample should be free of any {\em systematic} biases, which may 
not have been true for earlier analyses that typically used larger 
compilations of detected planets. For illustrative purposes, we 
divide the planets in our subsample into four logarithmic bins 
in semi-major axis (between 0.1~AU and 2.5~AU) and plot the number 
of planets in each bin over the theoretical curves in Figure~\ref{f5}. 
The observed distribution, in agreement with the theoretical models, 
rises rapidly with increasing semi-major axis.

\begin{figure}
\plotone{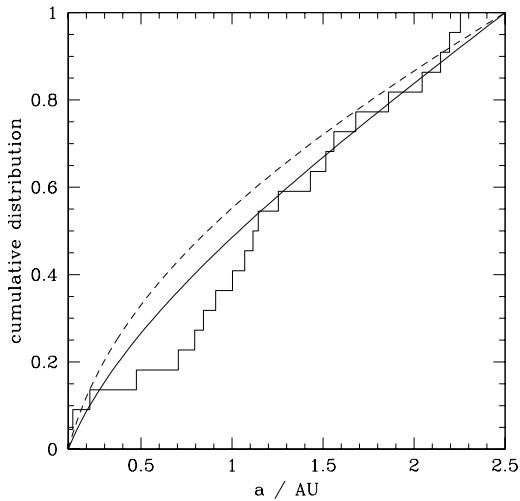}
\caption{Comparison of the predicted planet distribution with the observed one. 
The solid histogram shows the cumulative radial distribution of massive planets 
with $M_p \sin(i) > 1.65 M_J$. The sample is based on that of \cite{fischer05}, 
with the planet properties updated using the compilation of \cite{butler06}. The solid 
curve is the prediction of the fiducial model for planets of mass $3 M_J$ 
formed at 5~AU in a disk with $\gamma=1$ and partial suppression of the migration 
velocity. The dashed curve shows the predictions of a model in which $\gamma=1.5$ 
but all other parameters remain fixed. This, and other alternate models   
which predict flatter distributions, are not supported by the data, though the 
distributions cannot be distinguished at much better than the 95\% confidence level.}
\label{f6}
\end{figure}

Figure~\ref{f6} shows the cumulative distribution of orbital radii 
for massive planets in the sample, together with selected theoretical 
curves derived from the differential distributions plotted in Figure~\ref{f5}. 
The baseline model, in which planets form at a constant rate at relatively 
small orbital radii (5~AU) in a protoplanetary disk with $\gamma=1$, is shown as 
the solid line. This model is consistent with the available data 
(Kolmogorov-Smirnov test $P=0.3$). Of the variant models considered, 
those in which migration is fully suppressed, $\gamma > 1$, and / or planets are 
considered to typically form at larger radii are disfavored, as these 
models predict relatively more planets at sub-AU orbital radii that are 
not observed. A model in which $\gamma=1.5$, with the other parameters 
remaining unchanged, is shown as the upper  
dashed line. This model is disfavored at modest statistical significance 
(Kolmogorov-Smirnov test $P=0.06$). Several variant models that we have 
considered are disfavored at roughly comparable confidence levels 
(95\%), but with such a small observed sample few definitive conclusions 
can be drawn.

As noted above, our sample selection procedure is deliberately conservative. 
We have also compared the theoretical model to a larger sample, obtained by 
taking the entire catalog of extrasolar planets from \cite{butler06} and 
retaining those planets with orbital radii $0.1 \ {\rm AU} < a < 2.5 \ {\rm AU}$ 
and masses $1.65M_J < M_p \sin(i) < 10 M_J$ (i.e. the same mass and radius 
cuts as before, but relaxing the condition that all planets were detected 
from one survey with well-known selection properties). The resulting 
sample of 53 planets has a radial distribution 
that, by eye, looks very similar to the 22 planet sample, and the greater 
numbers give more power to discriminate between models. The baseline model 
($\gamma=1$, formation at 5~AU, partially suppressed migration) remains 
consistent with the data, while the second model discussed above in 
which $\gamma=1.5$ is more clearly inconsistent ($P=3 \times 10^{-3}$). 
We do not place much weight on these results, since the compilation of 
planets reported in \cite{butler06} is not uniformly selected. However, 
they illustrate clearly that only modest expansion of the 
sample size --- by around a factor of two --- would allow us (within the 
formal confines of the model) to test whether the observed distribution 
of massive extrasolar planets is or is not consistent with theoretical 
models that differ in their assumptions as to planet formation radius, 
migration rate, and disk properties. Since some of these parameters 
are partially degenerate, somewhat larger unbiased samples would be 
needed to try and pin down empirically a single favored model.

\section{Radial variation of the mass function}
At small orbital radii where $B < 1$, the migration velocity is 
dependent on the planet mass. In the case in which migration is 
partially suppressed, $\dot{r}_p \propto M_p^{-1/2}$, and hence the 
fraction of time over which the most massive planets ($10 M_p$) can 
form and survive without migrating in to the star is a factor of a few 
greater than for Jupiter mass planets. In a steady-state disk, 
however, the fractional suppression of the migration velocity as a function 
of radius is fixed for each planet mass (at radii where $B < 1$). 
Hence, it is approximately true that the 
{\em relative} number of planets left stranded at different radii 
is independent of the planet mass, and the predicted mass function 
of giant planets is independent of radius \citep{armitage02}.

\begin{figure}
\plotone{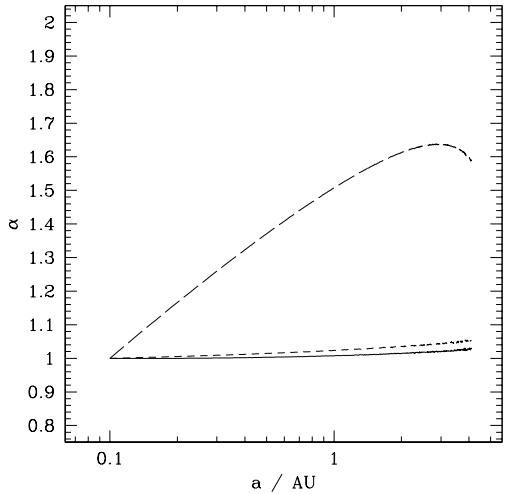}
\caption{The magnitude of the predicted variation in the extrasolar planet mass 
function with radius in different models. The solid curve shows results from the 
fiducial model in which planets form at 5~AU, $\gamma=1$, and the migration 
velocity is partially suppressed. The short-dashed curve shows results for $\gamma=1.5$, 
with the remaining model parameters unchanged from their fiducial values.  
The long dashed curve shows the effect of assuming that a 
change of migration regime occurs between high and low mass planets. In all 
cases the slope of mass function at small radii has been fixed at $\alpha=1$.}
\label{f7}
\end{figure}

In more detail, however, migration does lead to some fractionation 
of the planet mass function. More massive planets must form earlier 
than their less massive counterparts if they are to end up at the 
same orbital radii after disk dispersal, and the evolution of the 
disk is itself time-dependent (being dominated by viscous evolution 
early on, with a change to a steeper decline in the mass accretion 
rate later on due to the influence of photoevaporation). To quantify 
the extent to which this results in a radial variation of the 
mass function we assume that at small orbital radii the mass function 
can be written as a power-law,
\begin{equation}
 \frac{{\rm d}N_p}{{\rm d}M_p} \propto M_p^{-\alpha}
\end{equation}
and compute the predicted radial variation in $\alpha$. Figure~\ref{f7} 
shows the results for two of the models that we have discussed 
earlier, including the baseline model. For definiteness, we set $\alpha=1$ 
at small radii, though this choice is arbitrary. The predicted variation  
in the mass function with orbital radius in these models is non-zero, 
but in these and other similar models we have computed in 
which the migration properties of low and high mass planets are the same the 
magnitude of the change is small --- of the order of $\Delta \alpha = 0.1$ or less. 
It is straightforward to show that variations of this magnitude are 
undetectable with feasible surveys.  
The conclusion is that any detectable change in the mass function with 
radius (at small orbital radii where migration rather than mass growth 
is the dominant effect) would have to be attributed to effects other 
than mass-dependent migration.

There is one exception to this rule. If the migration {\em regime} 
for high mass planets is different from that for low mass planets, then 
large changes in the mass function with radius result. Such a change in 
migration regime is conceivable --- for example it is possible that high mass planets 
completely prevent mass flow across the gap, while low mass planets 
allow significant mass flow \citep{lubow99}. In this case, one would 
expect less perturbation to the disk structure immediately outside the 
gap for low mass planets that for high mass planets. If less mass 
piles up outside the tidal barrier, the torque will be reduced, 
and the migration velocity will go down. An example model of this kind, in which we assume 
that migration is fully suppressed for $0.5 M_J$ planets but only 
partially suppressed for $2 M_J$ planets, is also plotted in Figure~\ref{f7}. 
As expected (given the substantial difference between the relevant curves 
in Figure~\ref{f5}), this model displays order unity variations in the 
predicted slope of the mass function with radius. 

\begin{figure}
\plotone{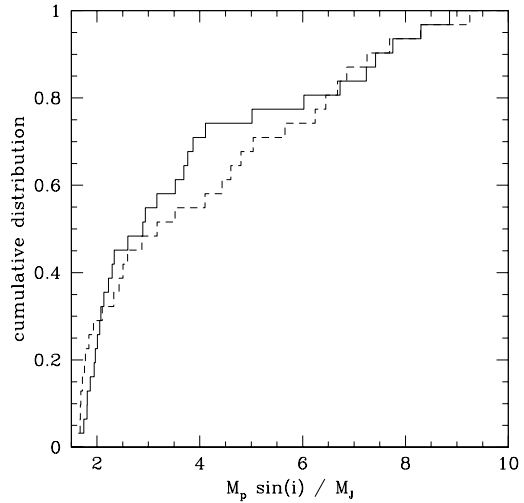}
\caption{The cumulative mass function of known extrasolar planets in the mass 
range $1.65 \ M_J < M_p \sin(i) < 10 \ M_J$ is plotted for two intervals in 
semi-major axis. The solid histogram shows the distribution for planets 
with $0.1 \ {\rm AU} < a < 1.185 \ {\rm AU}$, while the dashed histogram 
represents the distribution for $1.185 \ {\rm AU} < a < 2.5 \ {\rm AU}$.  
These distributions are not statistically distinguishable.}
\label{f7b}
\end{figure}

Observationally, an apparent paucity of high mass planets amongst the 
hot Jupiters has been noted by several authors \citep{zucker02}. These 
planets have presumably undergone interactions with their host stars 
(possibly including tidal circularization, mass loss, or stopping 
of migration due to entry into the stellar magnetosphere), any one of 
which could result in either mass loss or a mass-dependence to the 
planet's survival probability. At larger radii, however, there is no 
strong evidence for any variation in the mass function \citep{marcy05}. 
To illustrate this with current data, we consider the entire sample of 
known extrasolar planets \citep{butler06} with masses in the range 
$1.65M_J < M_p \sin(i) < 10 M_J$, and orbital radii between 
$0.1 \ {\rm AU} < a < 2.5 \ {\rm AU}$ (similar conclusions follow 
from analyses of smaller, more strictly selected samples). We divide this sample 
into `inner' and `outer' subsets, with equal numbers of planets 
in each. The dividing line between the subsets falls at a semi-major 
axis of 1.185~AU. Figure~\ref{f7b} shows the cumulative distributions  
of $M_p \sin(i)$ for these subsamples. By eye, and statistically, no 
significant differences are seen. This is consistent with the 
predictions of the simple migration model we have developed here, 
and probably already rules out the large changes in the mass function 
with radius predicted in the case where high and low mass planets 
migrate in a qualitatively distinct manner.

\section{Sensitivity to structure in the protoplanetary disk}
The structure of the protoplanetary disk at radii of the order of 
1~AU is uncertain theoretically, primarily as a consequence of the 
low ionization fraction which sows doubt as to the efficiency of 
angular momentum transport driven by the magnetorotational instability 
\citep{balbus98,gammie96,turner07}. Moreover, most existing observational constraints 
are either explicitly \citep{wilner00} or implicitly \citep{hartmann98,armitage03} 
sensitive only to large radii in the disk. Not only is the slope of the 
steady-state surface density profile at small radii ($\Sigma \propto r^{-\gamma}$) 
poorly known, but there could in principle be discontinuities in 
$\Sigma$ due to abrupt changes in the angular momentum transport 
efficiency \citep{gammie96}. We have looked at how both of these 
possibilities affect the resulting distribution of planets.

\begin{figure}
\plotone{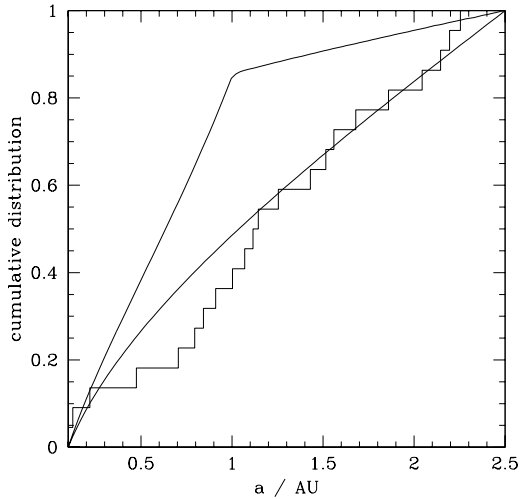}
\caption{Illustration of the effect of discontinuities in the efficiency of disk 
angular momentum transport on the predicted planet distribution. In this, not 
very realistic example, the viscosity is assumed to drop by a factor of 10 
inside 1~AU (upper curve), leading to a large accumulation of planets at small 
orbital radii. As in Figure~6, the histogram shows the observed planet 
distribution and the lower curve the prediction of the fiducial migration model.}
\label{f8}
\end{figure}

Holding other parameters of the model (primarily the planet formation 
radius) fixed, the resulting differential distributions of massive 
planets for different $\gamma$ are plotted in Figure~\ref{f5}. Here 
we have assumed a smooth surface density profile, which is fully 
characterized by the value of $\gamma$ which we vary between 
$\gamma=0.5$ and $\gamma=1.5$. The differences between these 
models at sub-AU radii are quite significant. If we consider the 
relative number of planets at 1~AU and 0.1~AU, the ratio varies 
from 2.4 ($\gamma=1.5$) to 3.8 ($\gamma=1$) to 6.6 ($\gamma=0.5$). 
The unbiased subsample defined earlier includes only a handful 
of planets at sub-AU radii, so comparison with existing data is not possible, 
but these differences are large enough that only a small unbiased 
sample of planets at these small radii would be needed to see 
variations of this magnitude. Changing the assumed planet formation 
radius has only a small influence on the distribution at sub-AU 
scales, so the influence of different $\gamma$ is in principle 
separable from other unknown parameters of the model.

Abrupt changes in the surface density profile --- for example 
due to opacity transitions within the disk or rapid changes in the 
efficiency of angular momentum transport with radius, leave an 
even more distinctive fingerprint in the radial distribution of 
planets, provided only that migration is not fully suppressed,  
so that $\dot{r}_p$ is not linear in $B$. Such changes lead to 
a jump in the differential distribution of planets at the radius 
where the discontinuity occurs, and correspond to a change of 
slope in the cumulative distribution. As an example, Figure~\ref{f8} 
shows the predicted cumulative distribution if the viscosity is assumed to 
drop by a factor of 10 within 1~AU. Only rather unrealistic toy models 
of this kind (in which the jump is both large and situated squarely 
in the middle of the accessible radial range) can be ruled out using 
existing data, but larger data sets could potentially constrain 
discontinuities in the disk physics quite well.

\section{Conclusions}
In this paper we have investigated the predictions of a rather 
simple, almost entirely analytic, model of giant planet formation 
and migration for the radial distribution of massive extrasolar 
planets. Our main results are:
\begin{enumerate}
\item
The distribution of massive planets with radius, inside the 
snow line beyond which planets are assumed to form, preserves 
information about the typical location of planet formation and 
the structure of the protoplanetary disk at small radii. Statistical 
analyses of large, uniformly selected samples of massive planets  
could therefore in principle be used to recover information about planet 
formation and disk physics.
\item
Abrupt changes in the disk properties with radius, which may occur 
at small orbital radii, leave the most distinctive signature in the 
resulting planet distribution (a corresponding discontinuity in 
${\rm d}N_p / {\rm d}\log a$). Constraints on smooth surface density 
profiles at AU scales are also possible, but are likely to harder 
to obtain and more ambiguous.
\item
The mass function of extrasolar planets is not predicted to vary 
detectably with radius as a consequence of migration, at least in the 
simplest models.
\item
Existing data remains compatible with a surprisingly simple model, 
proposed previously \citep{armitage02}, in which massive planets form beyond the snow line 
at a constant rate, and migrate inward through a smooth protoplanetary disk before 
becoming stranded when the gas disk is dispersed due to photoevaporation. 
However, the sample of planets suitable for a statistical comparison 
with the theory is limited, and as a result a wide class of 
alternative models remain viable or can only be ruled out with 
limited significance.
\end{enumerate}
Without a doubt, the simple theoretical model we have explored here is 
too simple. On a technical level, it would be desirable to extend the 
model to treat properly the case when planets form within a disk 
whose mass is only a small multiple of the planet mass, as this 
situation is a common outcome of the scenarios we favor. The model 
also ignores some effects, such as mass flow across the gap and 
accretion onto the planet, that are almost certainly important 
\citep{lubow99,lubow06}, along with others, such as planet-planet 
scattering which has the side-effect of altering the orbital 
elements of surviving planets \citep{ford01}, that could be 
significant. As such, the current agreement between our 
predictions and the observed distribution of extrasolar planets speaks 
more to the paucity of the data than to the validity of the theoretical model. 
Taking a broader view, however, the effects that we have ignored do not 
appear to be so intractable that they could not, in the future, be 
quantified and incorporated into a theoretical model of migration. Our results 
therefore imply that comparison 
of theoretical models to the larger samples of extrasolar planets that are 
forthcoming holds substantial promise for learning new information,  
about both planet formation and protoplanetary disk physics, that is 
currently unavailable via other methods.

\acknowledgements

I thank Richard Alexander for a detailed critique of an early version 
of this paper, and in particular for alerting me to the work of \cite{ruden04}. I 
am also grateful to the referee for an insightful report. 
This work was supported by NASA under grants NAG5-13207, NNG04GL01G and NNX07AH08G 
from the Origins of Solar Systems and Astrophysics Theory Programs, and 
by the NSF under grant AST~0407040.

\end{document}